\documentclass[aps,prb,twocolumn,noshowpacs]{revtex4-1}
\usepackage{amsmath,graphicx,amsbsy,amssymb,amsmath,epsfig,latexsym,wasysym,pifont,mathrsfs,natbib}
\usepackage{float} \newfloat{widefig}{thp}{lop} \usepackage{comment}
\usepackage{array, makecell} \usepackage{verbatim} \usepackage{datetime}
\usepackage{multirow,array} \usepackage{hyperref} \usepackage{color} \usepackage{setspace}

\begin{document}

\title{Understanding the origin of the magneto-caloric effects in substitutional Ni-Mn-Sb-Z (Z=Fe, Co, Cu) compounds: insights from first-principles calculations} 
\author{Sheuly Ghosh}
\email[]{sheuly.ghosh@iitg.ac.in}
\affiliation{Department of Physics, Indian Institute of Technology
  Guwahati, Guwahati-781039, Assam, India.}    
\author{Subhradip Ghosh}
\email{subhra@iitg.ac.in} \affiliation{Department of Physics,
  Indian Institute of Technology Guwahati, Guwahati-781039, Assam,
  India.} 

\begin{abstract}
Ni-Mn based ternary Heusler compounds have drawn attentions lately as significant magneto-caloric effects in some of them have been observed. Substitution of Ni and Mn by other $3d$ transition metals in controlled quantity have turned out to be successful in enhancing the effect and bring the operational temperatures closer to the room temperature. Using density functional theory calculations, in this work, we have systematically explored the roles of various factors such as site occupancies, magnetic interactions, and compositions associated with the constituents of Mn-excess Ni$_{2}$MnSb Heusler compounds upon substitution of Ni and/or Mn by $3d$ transition metals Fe, Co and Cu. Our calculations unveiled the physics behind the variations of physical properties associated with the magneto-caloric effects, and thus interpreted the available experimental results successfully. The work also provided important information on the compounds and the composition ranges where significant magneto-caloric effects may be realised and further experimental investigations need to be done. 
\end{abstract}


\maketitle


\section{Introduction}\label{introduction}
In recent years, magneto-caloric effect (MCE), the driving force for magnetic refrigeration technology, led to intensive research due to its superiority over the conventional gas refrigeration technology on account of energy  and environmental concerns. Gd has been considered as a benchmark material in magnetic refrigerator due to the discovery of significant magneto-caloric effect in it, as an outcome of a second-order ferromagnetic to paramagnetic transition close to room temperature \cite{brown1976}. Giant MCE is generally observed in rare-earth materials which undergo a magneto-structural transition and/or second order magnetic transition\cite{pecharsky2001,tishin2016,de2006,fujieda2002,wada2001,tegus2002}. \\
In this context, Ni-Mn based shape memory Heusler alloys drew considerable attention due to the presence of the martensitic phase transition (MPT), which in many cases is coupled with the magnetic transition, giving rise to a first order magneto-structural transition from the high temperature austenite (cubic) phase to the low temperature martensite (orthorhombic or tetragonal) phase\cite{li2012,li2014,krenke2005,7krenke2007}. This class of materials are less expensive and more effective than conventional rare-earth-based magnetocaloric materials. Several materials in this family have shown substantial MCE, thus raising their stakes as commercially viable materials for magneto-caloric applications. In the Ni-Mn Heusler family, apart from Ni$_{2.2}$Mn$_{0.8}$Ga which exhibits the highest MCE\cite{8pasquale2005}, off-stoichiometric Ni-Mn-Sn and Ni-Mn-In systems with Mn-excess and In/Sn deficient compositions exhibit substantial MCE in the vicinity of first-order magnetostructural transition \cite{pathak2007,79krenke2005,muthu2010}. All these studies on magnetocaloric materials conclude that from the perspective of practical applications, high-performance magnetocaloric materials should meet at least the following important requirements:  (1) the materials should posses a good value of magnetisation in the high temperature phase. A large difference in magnetisation ($\Delta$M) between the high temperature austenite and low temperature martensitic phases is also of great importance because  substantial Zeeman energy, which is defined as $\Delta$M.H, is the key to first order magneto-structural transition; (2) the first-order magneto-structural transition temperature (T$_{M}$) and the second-order magnetic transition temperature (T$_c^A$) in the austenite phase should be close and must be near room temperature; (3) the materials should have good mechanical stability during operation; (4) the two parameters characterising the MCE, the magnetic-field-induced isothermal entropy change ($\Delta$S$_{M}$) and adiabatic temperature change($\Delta$T$_{ad}$), must be significant at a magnetic field as low as possible. \\

Inspite of the discovery of substantial MCE in Mn-excess, Z-deficient off-stoichiometric Ni-Mn-Z Heusler alloys some serious drawbacks  due to the off-stoichiometry limited their potentials. One major disadvantage is that the value of $\Delta$M is limited due to the low value of magnetisation in the high temperature austenite phase which arises because of the antiparallel alignment between the Mn atoms occupying different crystallographic sites. To overcome this, recently, several experimental studies have been done on Ni-Mn-Z compounds  considering substitution of either of the constituents with another 3$d$-transition element (i) to get a larger $\Delta$M near the structural transition (ii) to tune the transition temperatures (T$_{M}$ and T$_c^A$) and bring them close to each other, (iii) to improve the mechanical properties. Positive changes in several counts were observed for substitution of Co, Cu and Fe at different sites and in different proportions \cite{agao2009,lee2009,12kainuma2006,khan2005,soto2010,pathak2010,gomes2006,stadler2006,soto2008,cherechukin2001,wang2006,jing2009,krenke2007,gao2009}.

A recent addition to this class of Heusler alloys showing promising MCE properties, is the off-stoichiometric Mn-excess Sb-deficient Ni-Mn-Sb systems where magneto-structural transition and significant magneto-caloric effect are observed near room temperature \cite{91duc2012,22khan2007}. This, along with low cost of Sb and achievable negligible hysteresis loss \cite{20w2009}, makes Ni-Mn-Sb systems of great interest in the research on magnetic refrigeration. A large positive  $\Delta$S$_{M}$ of 19~Jkg$^{-1}$K$^{-1}$ at 297~K was obtained under a field of 5~T in Ni$_{2}$Mn$_{1.48+x}$Sb$_{0.52-x}$ with $x=0.04$\cite{22khan2007}. With an aim to improve the MCE, some recent investigations have also been carried out for the Mn-excess Sb-deficient off-stoichiometric Ni-Mn-Sb systems by substituting the Fe and Co atoms at Mn and Ni sites respectively. As a consequence, a large positive $\Delta$S$_{M}$ near room temperature was obtained with 0.28$\leq$x$\leq$0.36 in Ni$_{2-x}$Co$_{x}$Mn$_{1.56}$Sb$_{0.44}$ alloys\cite{64han2008}. For a slightly different composition, Ni$_{2-x}$Co$_{x}$Mn$_{1.52}$Sb$_{0.48}$, nearly 70\% decrease in moment is observed associated with the martensitic transition and remarkable enhancement in $\Delta$S$_{M}$ of 34~Jkg$^{-1}$K$^{-1}$ is achieved for $x=0.2$ at 262~K in a field of 5~T near room temperature\cite{nayak2009,anayak2009}. A significant $\Delta$S$_{M}$ value of 14.2~Jkg$^{-1}$K$^{-1}$ at 288~K under 5~T field was observed for $x=0.08$ in Ni$_{2}$Mn$_{1.52-x}$Fe$_{x}$Sb$_{0.48}$\cite{59sahoo2011}. Thus, the Ni-Mn-Sb based Heusler alloys show a possibility to be proved as emerging materials, showing significant MCE properties with the substitution of $3d$-transition elements. 

These results suggest that Fe, Co and Cu substitutions at select sites of Ni-Mn-Z Heusler compounds help in improving their MCE properties. However, it crucially depends on the substituent, the site of substitution and the composition. What is nevertheless lacking is a systematic investigation into the impacts of these three. Such an investigation would throw light on the microscopic understanding of these factors, help interpret the experimental observations and provide a roadmap to tune the compositions for maximising the functional properties. in this work, we have considered relatively less explored Ni-Mn-Sb system to address these issues. We specifically consider the compound Ni$_{2}$Mn$_{1.5}$Sb$_{0.5}$, relative concentrations of the elements being around the mostly studied experimental composition, as the parent one, and systematically substitute Ni and Mn by Fe, Co and Cu, varying the concentrations of the substituents. We mainly address the following: (i) the effect on the magnetisation in the high temperature phase and whether and how the substitutions improve $\Delta$M (ii) in what way the substitutions impact the structural phase stabilities, the magnetic exchange interactions, the mechanical properties, the characteristic temperatures (T$_{M}$ and T$_c^A$) and (iii) how the information from (i) and (ii) can be correlated to the magneto-caloric properties of the Ni-Mn-Sb compounds. In light of these, we also attempt to interpret the experimental observations in Fe-substituted\cite{59sahoo2011} and Co-substituted\cite{64han2008,nayak2009,anayak2009} compounds which highlight the important role of atomic ordering in the system.


\section{Details of calculations}\label{methods}
At high temperature, Ni$_2$MnSb crystallises in a Cu$_2$MnAl-type structure (regular Heusler L2$_1$, space group no. 225 (Fm$\bar{3}$m))\cite{72khan2008,20w2009,53rama2011} with three inequivalent Wyckoff positions (4a, 4b, 8c). The Sb  and Mn atoms occupy 4a (0, 0, 0) and 4b (0.50, 0.50, 0.50) positions respectively and Ni atoms occupy the 8c ((0.25, 0.25, 0.25) and (0.75, 0.75, 0.75)) sites. In the present work, we have dealt with Fe, Co and Cu substituted Ni$_{2}$Mn$_{1.50}$Sb$_{0.50}$ compounds with substitutions done both at Ni and Mn sites to get a comparative  understanding of how they affect the properties as a function of composition. We, thus, have considered two different systems: (i) Ni$_{2}$Mn$_{1.50-y}$Z$_{y}$Sb$_{0.50}$(referred as Z@Mn) and (ii) Ni$_{2-y}$Z$_{y}$Mn$_{1.50}$Sb$_{0.50}$(referred as Z@Ni) with Z=Fe, Co and Cu for $y=0, 0.25, 0.50$ (shown in Table \ref{table1}). The systems are modelled with a 16-atom conventional cubic cell. For example, to make a 25\% Fe-substituted Ni$_2$Mn$_{1.25}$Fe$_{0.25}$Sb$_{0.50}$ composition, one Mn atom out of the six in the conventional cell of Ni$_2$Mn$_{1.50}$Sb$_{0.50}$ is replaced with one Fe atom, as done elsewhere\cite{54v2015,55sokolovskiy2015,28kundu2017,ghosh2019}. Though the experiments have been done with  much smaller concentrations of the substituents, we had to restrict ourselves with above mentioned values of $y$ as modelling of experimental compositions would require computationally demanding larger supercells. 

Electronic structure calculations were done with spin-polarised density functional theory (DFT) based projector augmented wave (PAW) method as implemented in Vienna {\it Ab initio} Simulation Package (VASP)\cite{41blochl1994,43kresse1999,42kresse1996}. The valence electronic configurations used for the Mn, Fe, Co, Ni, Cu and Sb PAW pseudopotentials are 3$d^{6}$4$s$, 3$d^{7}$4$s$, 3$d^{8}$4$s$, 3$d^{8}$4$s^{2}$, 3$d^{10}$4$s$ and 5$s^{2}$5$p^{3}$, respectively. For all calculations, we have used the Perdew-Burke-Ernzerhof implementation of generalised gradient approximation for exchange-correlation functional\cite{44perdew1996}. An energy cut off of 550 eV and a Monkhorst-Pack $11\times{11}\times{11}$ k-mesh were used for self-consistent calculations. The convergence criteria for the total energies and the forces on individual atoms were set to 10$^{-6}$ eV and $10^{-2}$ eV/\r{A} respectively for all calculations. 

To study the variation of Curie temperatures with concentrations, we have calculated the magnetic pair exchange parameters using multiple scattering Green function formalism(KKR) as implemented in SPRKKR code\cite{ebert2011}. In here, the spin part of the Hamiltonian is mapped to a Heisenberg model

\begin{eqnarray}
H = -\sum_{\mu,\nu}\sum_{i,j}
J^{\mu\nu}_{ij}
\mathbf{e}^{\mu}_{i}
.\mathbf{e}^{\nu}_{j}
\end{eqnarray}

$\mu$, $\nu$ represent different sub-lattices, \emph{i}, \emph{j} represent atomic positions and $\mathbf{e}^{\mu}_{i}$ denotes the unit vector along the direction of magnetic moments at site \emph{i} belonging to sub-lattice $\mu$. The $J^{\mu \nu}_{ij}$s are calculated from the energy differences due to infinitesimally small orientations of a pair of spins within the formulation of Liechtenstein { \it et al.}\cite{liechtenstein1987}. In order to calculate the energy differences by the SPRKKR code, full potential spin polarized scalar relativistic Hamiltonian with angular momentum cut-off $l_{max} = 3$ is used along with a converged k-mesh for Brillouin zone integrations. The Green's functions were calculated for 32 complex
energy points distributed on a semi-circular contour. The energy convergence criterion was set to 10$^{-5}$ eV for the self-consistent cycles. Here we used the equilibrium lattice parameters and the optimized atomic positions from the projector augmented wave calculation using VASP package. These exchange parameters are then used for the calculations of Curie temperatures (T$_c^A$). The Curie temperatures are estimated with two different approaches: the mean-field approximation (MFA)\cite{26sokolovskiy2012,meinert2010} and the Monte Carlo simulation (MCS) method\cite{landau2014,zagrebin2016} in order to check the qualitative consistency in the results and to obtain a reliable estimate of the quantity as the MFA is known to overestimate T$_{c}^A$ while the MCS method is more accurate quantitatively. Details of the MFA and MCS calculations are given in supplementary material. \\

The elastic constants for the compounds are calculated using energy-strain method only for high-temperature cubic austenite phases. To determine the bulk modulus (B), the total energy vs volume data is fitted to $Murnaghan's$ equation\cite{46murnaghan1944}. Then the elastic moduli C$^\prime$ and C$_{44}$ are calculated\cite{47vitos2007,48kart2010} by considering volume conserving orthorhombic($\epsilon_o$) and monoclinic($\epsilon_m$) deformations  of the cubic cell, respectively. Six strains $\epsilon$=0, 0.01, 0.02, 0.03, 0.04, 0.05 were used to calculate the total energies E($\epsilon_o$) and E($\epsilon_m$). The elastic moduli (C$^\prime$ and C$_{44}$) are then obtained by fitting the variation of total energies with distortions to a fourth order polynomial equation\cite{48kart2010}. C$_{11}$ and C$_{12}$ are then calculated using the relations: B=$\frac{1}{3}$(C$_{11}$+2C$_{12}$) and C$^\prime$=$\frac{1}{2}$(C$_{11}$-C$_{12}$). The isotropic shear modulus, G is typically calculated as an average of G$_v$, according to the formalism of Voigt\cite{49voigt1889} and G$_R$, according to the formalism by Reuss\cite{50roy2015}. In cases of a number of ferromagnetic Heusler alloys, it was found out that G$_v$ using Voigt formalism are closer to the experimental results\cite{26sokolovskiy2012,51roy2016}. Hence we have approximated G as G$_v$ and calculated its value using the relation: G$_v$=$\frac{1}{5}$(C$_{11}$-C$_{12}$+3C$_{44}$). Finally Cauchy pressure (C$^p$) has been calculated as C$^p$=(C$_{12}$-C$_{44}$).


\section{Results and Discussions}\label{results}
\subsection{Site Preferences, and magnetic ground states in austenite phases} \label{table1}

The configurational ordering of the constituent elements in the crystal lattice affects both the martensitic transformation characteristics and the magnetic properties of Ni-Mn-based alloys \cite{26sokolovskiy2012,ghosh2019,82ghosh2014,35sanchez2007}. First-principles calculations\cite{45li2011,28bkundu2017,89chakrabarti2013} demonstrated that the substituent $3d$-transition metals do not always prefer to occupy the  sites of substitutions. Therefore, we first focus on determination of the site preferences and the associated magnetic ground states of the substituted Ni$_{2}$Mn$_{1.5}$Sb$_{0.5}$ compounds in their high temperature austenite phases, by comparing total energies of various possible site ordered and magnetic configurations at fixed compositions. The results are shown in Table ~\ref{table1}.

\begin{table*}[htpb!]
\centering
\caption{\label{table1} Preferred site-occupancies, corresponding possible magnetic configurations and their relative electronic energies $\Delta$E$_{0}$ (in meV/atom) are shown. $``$C1$"$ to $``$C4$"$ denote possible magnetic configurations, for Z=Fe, Co and Cu substituted (i) Ni$_{2}$Mn$_{1.50-y}$Z$_{y}$Sb$_{0.50}$(Z@Mn) and (ii) Ni$_{2-y}$Z$_{y}$Mn$_{1.50}$Sb$_{0.50}$(Z@Ni) systems. The atom $``$X$"$  at its original site in L2$_{1}$ lattice is denoted as $``$X1$"$, whereas it is denoted as $``$X2$"$ when it occupies any other site. The reference energy in each case is that of $``$C1$"$ or $``$C3$"$ (when $``$C1$"$ is not possible) magnetic configuration. Boldface indicates the ground state magnetic configuration for the corresponding composition.}  
\resizebox{0.93\textwidth}{!}{%
\vspace{0.3 cm}
\begin{tabular}{l@{\hspace{0.3cm}} c@{\hspace{0.2cm}} c@{\hspace{0.2cm}} c@{\hspace{0.2cm}} l@{\hspace{0.2cm}} c@{\hspace{0.2cm}} }
\hline\hline
\vspace{-0.33 cm}
\\(i) Ni$_{2}$Mn$_{1.50-y}$Z$_{y}$Sb$_{0.50}$ \\
\hline\hline
\vspace{-0.33 cm}
\\ Composition & \multicolumn{3}{c}{Site Occupancies} & Mag. Configurations & $\Delta$E$_{0}$ \\
               &    4a   &   4b   &    8c             &                     &       \\
\hline\hline
\\Z=Fe \\
\hline
Ni$_{2}$Mn$_{1.50}$Sb$_{0.50}$(y=0.00) & Sb1$_{0.50}$Mn2$_{0.50}$ & Mn1 & Ni1$_{2}$ & \textbf{C1(Mn1$\uparrow$ Mn2$\downarrow$ Ni1$\uparrow$)} & 0.00 \\
                                       &                        &    &          & C2(Mn1$\uparrow$ Mn2$\uparrow$ Ni1$\uparrow$)                 & 15.32 \\
Ni$_{2}$Mn$_{1.25}$Fe$_{0.25}$Sb$_{0.50}$(y=0.25) & Sb1$_{0.50}$Mn2$_{0.50}$ & Mn1$_{0.75}$Fe1$_{0.25}$ & Ni1$_{2}$ & C1(Mn1$\uparrow$ Mn2$\downarrow$ Ni1$\uparrow$ Fe1$\downarrow$) & 0.00 \\
                                                  &                          &                          &                          & C2(Mn1$\uparrow$ Mn2$\uparrow$ Ni1$\uparrow$ Fe1$\downarrow$) & -3.15 \\
                                                  &                          &                          &                          & \textbf{C3(Mn1$\uparrow$ Mn2$\downarrow$ Ni1$\uparrow$ Fe1$\uparrow$)} & -26.10 \\
                                                  &                          &                          &                          & C4(Mn1$\uparrow$ Mn2$\uparrow$ Ni1$\uparrow$ Fe1$\uparrow$) & -1.20 \\

Ni$_{2}$MnFe$_{0.50}$Sb$_{0.50}$(y=0.50) & Sb1$_{0.50}$Mn2$_{0.50}$ & Mn1$_{0.50}$Fe1$_{0.50}$ & Ni1$_{2}$ & C1(Mn1$\uparrow$ Mn2$\downarrow$ Ni1$\uparrow$ Fe1$\downarrow$) & 0.00 \\
                                                  &                          &                          &                          & C2(Mn1$\uparrow$ Mn2$\uparrow$ Ni1$\uparrow$ Fe1$\downarrow$) & -16.94 \\
                                                  &                          &                          &                          & \textbf{C3(Mn1$\uparrow$ Mn2$\downarrow$ Ni1$\uparrow$ Fe1$\uparrow$)} & -36.17 \\
                                                  &                          &                          &                          & C4(Mn1$\uparrow$ Mn2$\uparrow$ Ni1$\uparrow$ Fe1$\uparrow$) & -3.34 \\ 
\hline
\\Z=Co \\
\hline
Ni$_{2}$Mn$_{1.50}$Sb$_{0.50}$(y=0.00) & Sb1$_{0.50}$Mn2$_{0.50}$ & Mn1 & Ni1$_{2}$ & \textbf{C1(Mn1$\uparrow$ Mn2$\downarrow$ Ni1$\uparrow$)} & 0.00 \\
                                       &                        &    &          & C2(Mn1$\uparrow$ Mn2$\uparrow$ Ni1$\uparrow$)  & 15.32 \\

Ni$_{2}$Mn$_{1.25}$Co$_{0.25}$Sb$_{0.50}$(y=0.25) & Sb1$_{0.50}$Mn2$_{0.50}$ & Mn1$_{0.75}$Ni2$_{0.25}$ & Ni1$_{1.75}$Co1$_{0.25}$ & \textbf{C3(Mn1$\uparrow$ Mn2$\downarrow$ Ni1,Ni2$\uparrow$ Co1$\uparrow$)} & 0.00 \\
                                                  &                          &                          &                          & C4(Mn1$\uparrow$ Mn2$\uparrow$ Ni1,Ni2$\uparrow$ Co1$\uparrow$) & 12.02 \\

Ni$_{2}$MnCo$_{0.50}$Sb$_{0.50}$(y=0.50) & Sb1$_{0.50}$Mn2$_{0.50}$ & Mn1$_{0.50}$Ni2$_{0.50}$ & Ni1$_{1.50}$Co1$_{0.50}$          & C3(Mn1$\uparrow$ Mn2$\downarrow$ Ni1,Ni2$\uparrow$ Co1$\uparrow$) & 0.00 \\
                                                  &                          &                          &                          & \textbf{C4(Mn1$\uparrow$ Mn2$\uparrow$ Ni1,Ni2$\uparrow$ Co1$\uparrow$)} & -0.69 \\
\hline
\\Z=Cu \\
\hline
Ni$_{2}$Mn$_{1.50}$Sb$_{0.50}$(y=0.00) & Sb1$_{0.50}$Mn2$_{0.50}$ & Mn1 & Ni1$_{2}$ & \textbf{C1(Mn1$\uparrow$ Mn2$\downarrow$ Ni1$\uparrow$)} & 0.00 \\
                                       &                        &    &          & C2(Mn1$\uparrow$ Mn2$\uparrow$ Ni1$\uparrow$)  & 15.32 \\
Ni$_{2}$Mn$_{1.25}$Cu$_{0.25}$Sb$_{0.50}$(y=0.25) & Sb1$_{0.50}$Mn2$_{0.25}$Cu1$_{0.25}$ & Mn1 & Ni1$_{2}$ & \textbf{C3(Mn1$\uparrow$ Mn2$\downarrow$ Ni1$\uparrow$ Cu1$\uparrow$)} & 0.00 \\
                                       &                        &    &          & C4(Mn1$\uparrow$ Mn2$\uparrow$ Ni1$\uparrow$ Cu1$\uparrow$) & 6.80   \\

Ni$_{2}$MnCu$_{0.50}$Sb$_{0.50}$(y=0.50) & Sb1$_{0.50}$Cu1$_{0.50}$ & Mn1 & Ni1$_{2}$ & \textbf{C4(Mn1$\uparrow$ Ni1$\uparrow$ Cu1$\uparrow$)} & 0.00 \\
\hline\hline
\\
\\

\hline\hline
\vspace{-0.33 cm}
\\(ii) Ni$_{2-y}$Z$_{y}$Mn$_{1.50}$Sb$_{0.50}$ \\
\hline\hline
\vspace{-0.33 cm}
\\ Composition & \multicolumn{3}{c}{Site Occupancies} & Mag. Configurations & $\Delta$E$_{0}$ \\
               &    4a   &   4b   &    8c   \\
\hline\hline
\\Z=Fe \\
\hline
Ni$_{2}$Mn$_{1.50}$Sb$_{0.50}$(y=0.00) & Sb1$_{0.50}$Mn2$_{0.50}$ & Mn1 & Ni1$_{2}$ & \textbf{C1(Mn1$\uparrow$ Mn2$\downarrow$ Ni1$\uparrow$)} & 0.00 \\
                                       &                        &    &          & C2(Mn1$\uparrow$ Mn2$\uparrow$ Ni1$\uparrow$) & 15.32  \\
Ni$_{1.75}$Fe$_{0.25}$Mn$_{1.50}$Sb$_{0.50}$(y=0.25) & Sb1$_{0.50}$Mn2$_{0.50}$ & Mn1 & Ni1$_{1.75}$Fe1$_{0.25}$ & C1(Mn1$\uparrow$ Mn2$\downarrow$ Ni1$\uparrow$ Fe1$\downarrow$) & 0.00 \\
                                                  &                          &                          &                          & C2(Mn1$\uparrow$ Mn2$\uparrow$ Ni1$\uparrow$ Fe1$\downarrow$) & 25.13 \\
                                                  &                          &                          &                          & \textbf{C3(Mn1$\uparrow$ Mn2$\downarrow$ Ni1$\uparrow$ Fe1$\uparrow$)} & -13.07 \\
                                                  &                          &                          &                          & C4(Mn1$\uparrow$ Mn2$\uparrow$ Ni1$\uparrow$ Fe1$\uparrow$) & -4.26 \\

Ni$_{1.50}$Fe$_{0.50}$Mn$_{1.50}$Sb$_{0.50}$(y=0.50) & Sb1$_{0.50}$Mn2$_{0.50}$ & Mn1 & Ni1$_{1.50}$Fe1$_{0.50}$ & C1(Mn1$\uparrow$ Mn2$\downarrow$ Ni1$\uparrow$ Fe1$\downarrow$) & 0.00 \\
                                                  &                          &                          &                          & C2(Mn1$\uparrow$ Mn2$\uparrow$ Ni1$\uparrow$ Fe1$\downarrow$) & 38.84 \\
                                                  &                          &                          &                          & \textbf{C3(Mn1$\uparrow$ Mn2$\downarrow$ Ni1$\uparrow$ Fe1$\uparrow$)} & -22.73 \\
                                                  &                          &                          &                          & C4(Mn1$\uparrow$ Mn2$\uparrow$ Ni1$\uparrow$ Fe1$\uparrow$)& -13.68 \\
\hline
\\Z=Co \\
\hline
Ni$_{2}$Mn$_{1.50}$Sb$_{0.50}$(y=0.00) & Sb1$_{0.50}$Mn2$_{0.50}$ & Mn1 & Ni1$_{2}$ & \textbf{C1(Mn1$\uparrow$ Mn2$\downarrow$ Ni1$\uparrow$)} & 0.00 \\
                                       &                        &    &          & C2(Mn1$\uparrow$ Mn2$\uparrow$ Ni1$\uparrow$) & 15.32  \\
Ni$_{1.75}$Co$_{0.25}$Mn$_{1.50}$Sb$_{0.50}$(y=0.25) & Sb1$_{0.50}$Mn2$_{0.50}$ & Mn1 & Ni1$_{1.75}$Co1$_{0.25}$ & \textbf{C3(Mn1$\uparrow$ Mn2$\downarrow$ Ni1$\uparrow$ Co1$\uparrow$)} & 0.00 \\
                                                  &                          &                          &                          & C4(Mn1$\uparrow$ Mn2$\uparrow$ Ni1$\uparrow$ Co1$\uparrow$) & 5.77 \\
Ni$_{1.50}$Co$_{0.50}$Mn$_{1.50}$Sb$_{0.50}$(y=0.50) & Sb1$_{0.50}$Mn2$_{0.50}$ & Mn1 & Ni1$_{1.50}$Co1$_{0.50}$ & C3(Mn1$\uparrow$ Mn2$\downarrow$ Ni1$\uparrow$ Co1$\uparrow$) & 0.00 \\
                                                  &                          &                          &                          & \textbf{C4(Mn1$\uparrow$ Mn2$\uparrow$ Ni1$\uparrow$ Co1$\uparrow$)} & -3.04 \\
\hline
\\Z=Cu \\
\hline
Ni$_{2}$Mn$_{1.50}$Sb$_{0.50}$(y=0.00) & Sb1$_{0.50}$Mn2$_{0.50}$ & Mn1 & Ni1$_{2}$ & \textbf{C1(Mn1$\uparrow$ Mn2$\downarrow$ Ni1$\uparrow$)} & 0.00 \\
                                       &                        &    &          & C2(Mn1$\uparrow$ Mn2$\uparrow$ Ni1$\uparrow$) & 15.32   \\
Ni$_{1.75}$Cu$_{0.25}$Mn$_{1.50}$Sb$_{0.50}$(y=0.25) & Sb1$_{0.50}$Mn2$_{0.50}$ & Mn1 & Ni1$_{1.75}$Cu1$_{0.25}$ & \textbf{C3(Mn1$\uparrow$ Mn2$\downarrow$ Ni1$\uparrow$ Cu1$\uparrow$)} & 0.00 \\
                                                  &                          &                          &                          & C4(Mn1$\uparrow$ Mn2$\uparrow$ Ni1$\uparrow$ Cu1$\uparrow$) & 17.74 \\
Ni$_{1.50}$Cu$_{0.50}$Mn$_{1.50}$Sb$_{0.50}$(y=0.50) & Sb1$_{0.50}$Mn2$_{0.50}$ & Mn1 & Ni1$_{1.50}$Cu1$_{0.50}$ & \textbf{C3(Mn1$\uparrow$ Mn2$\downarrow$ Ni1$\uparrow$ Cu1$\uparrow$)} & 0.00 \\
                                                  &                          &                          &                          & C4(Mn1$\uparrow$ Mn2$\uparrow$ Ni1$\uparrow$ Cu1$\uparrow$) & 18.90 \\
 
\hline\hline

\end{tabular}
}
\end{table*}

The results suggest the following:  Substitutions at Mn sites show that the different substituents prefer different sites. While the substituting Fe atoms prefer to occupy the Mn sublattices, Co atoms prefer the Ni sublattices forcing the Ni to occupy the substituted Mn sublattices. Substituting Cu atoms, on the other hand, prefer the  Sb sublattices. In case of substitutions at  Ni sites, all three substituents have preferences for the sites of substitution only, corroborating the experimental predictions in case of Co substituted  Ni$_{2}$Mn$_{1.52}$Sb$_{0.48}$\cite{60sahoo2014}. Regarding the preferred magnetic configurations, it can be concluded that for both Fe and Cu substitutions, the Mn1 atoms align parallel with Ni1, Ni2 and Z($=$Fe, Cu) atoms and align antiparallel with the Mn2 atoms; the configuration denoted as $``$C3$"$. In cases of Co substitutions, though the Mn atoms have anti-parallel alignments for smaller concentrations (i.e. $``$C3$"$ configuration), with  increase in Co concentration they align parallel making the $``$C4$"$ magnetic configuration as the ground state. However in Ni$_{2}$MnCo$_{0.50}$Sb$_{0.50}$ the energy difference between the $``$C3$"$ and $``$C4$"$ configurations is small indicating that a mixed phase of these two can be present. This implies that Co acts as a $``$ferromagnetic activator$"$ in Ni-Mn-Sb alloys as is seen in other Ni-Mn based Heusler alloys\cite{jing2009,vasiliev2010,han2010,perez2013,nayak2010}.

\subsection{Magnetic moments in the austenite phases} \label{moment}
\begin{figure}[t]
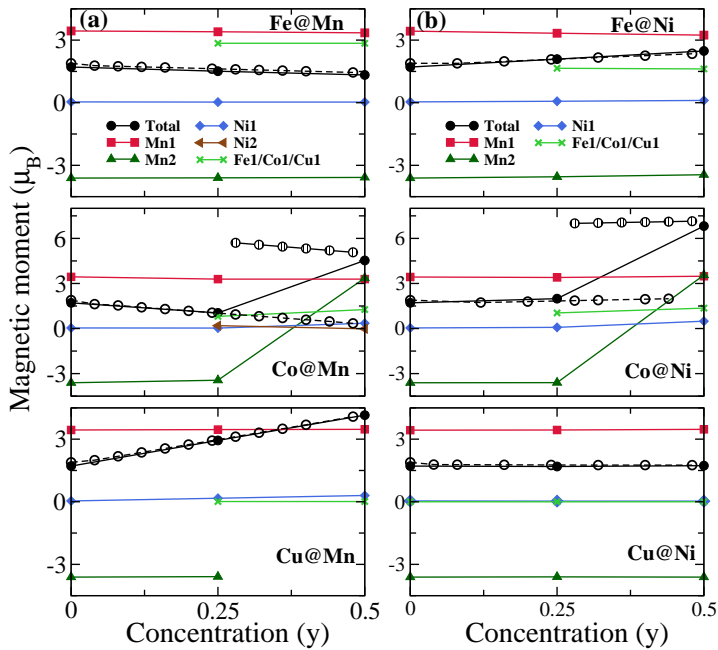

\centerline{\hfill
\psfig{file=fig1a.eps,width=0.245\textwidth}
\psfig{file=fig1b.eps,width=0.245\textwidth}
\hfill}
\caption{Calculated total magnetic moment and atomic magnetic moments (in $\mu_{\rm B}/f.u.$) (for atom-name convention see Table ~\ref{table1}) as a function of concentration $y$ of substituting elements, Z=Fe, Co and Cu for (a) Ni$_{2}$Mn$_{1.50-y}$Z$_{y}$Sb$_{0.50}$(Z@Mn) and (b) Ni$_{2-y}$Z$_{y}$Mn$_{1.50}$Sb$_{0.50}$(Z@Ni) systems in their lowest energy magnetic configurations as indicated in Table \ref{table1}. Variations of total magnetic moments with $y$ calculated by SPRKKR code\cite{ebert2011}, are presented by dashed lines with open symbols. For Co-substituted systems, magnetic moments both for $``$C3$"$(dashed lines with open circles) and $``$C4$"$(solid lines with marked circles) magnetic configurations for $y>0.25$ are shown.} 
\label{fig1}
\end{figure}

As was discussed in Section \ref{introduction}, significant enhancement of magnetic moment in the austenite phases leading to a possibility of large $\Delta$M is one of the motivations for substitutions of Ni and Mn with other magnetic atoms. That there is a correlation between enhancement of magnetisation in the austenite phase and a large MCE for Ni-Mn-Sb system could be observed in the experiment on the compound where Co substituted Ni. An enhancement in magnetisation in the austenite phase along with a significant MCE was observed\cite{nayak2009,anayak2009,nayak2010}. In order to understand the qualitative and quantitative trends in magnetisations due to substitutions of different elements at different sites and with different concentrations, we have calculated the total and atomic moments of all compounds as function of compositions. The results are shown in Table \ref{table2} and Figure \ref{fig1}. In Figure \ref{fig1}, we present simultaneously the variations in the moments when concentrations of the substituting element are between $0$ and $25 \%$ as well as between $25\%$ and $50\%$, that is when the concentrations could be chosen arbitrarily. These calculations are done by SPRKKR code which implements the KKR-CPA method that uses single-site mean field technique to address the substitutional disorder and hence do not require constructions of supercell. This was necessary in order to mimic the experimental compositions as much as possible and at the same time to find whether the qualitative changes in the moments with compositions indeed follow the trends as computed using supercells where concentrations of the substituents are varied by larger percentages. The lattice constants used for the calculations with KKR-CPA for arbitrary concentrations are taken by interpolating the lattice constants obtained from supercell calculations, presented in Table \ref{table2}.

We find that, Fe-substitutions at the  Mn1 sites in Ni$_{2}$Mn$_{1.50}$Sb$_{0.50}$ lead to a slight decrease in total magnetic moments with increase in the concentration of Fe, although the atomic moments hardly change. This is due to the lower moments on Fe atoms in comparison to the substituted Mn1 atoms. When Fe substitutes Ni, the total magnetic moment increases with Fe concentration as stronger magnet Fe replaces Ni. In case of Cu substituting Mn, we find remarkable increase in the total moment as Cu concentration increases. This occurs as the Cu atoms, instead of occupying the Mn1 sites, replace Mn2 atoms at the Sb sites, thus decreasing the negative contributions from Mn2 atoms to the overall moment. This does not happen when Cu replaces  Ni. Co-substitutions at either Mn or Ni sites present an interesting picture. When Co substitutes Mn or Ni, the magnetic configuration upto at least $y=0.25$, is $``$C3$"$. Since Co always occupies Ni sites irrespective of whether Mn or Ni is substituted, leaving Mn2 as it is, the moment decreases when Mn is substituted, as the net positive contribution to the total moment goes down with weaker magnet Co replacing Mn. When Ni is substituted, the net moment increases with $y$, albeit weakly, as Co moment is greater than that of Ni. Quantitatively the results of supercell calculations(by VASP) and KKR-CPA calculations (by SPRKKR) have excellent agreements, and reproduce experimental results well as KKR-CPA calculated moment value of 1.99 $\mu_{\rm B}/f.u.$ for Ni$_{1.8}$Co$_{0.2}$Mn$_{1.50}$Sb$_{0.50}$ is in good agreement with the experimental value of 1.85 $\mu_{\rm B}/f.u.$ for Ni$_{1.8}$Co$_{0.2}$Mn$_{1.52}$Sb$_{0.48}$ composition\cite{60sahoo2014}. For $y=0.5$, we found that although the lowest energy magnetic configuration $``$C4$"$, the energy difference between $``$C3$"$ and $``$C4$"$ is extremely low, even less than 1~meV per atom for Mn-substituted compound. This gives rise to the possibility of mixed magnetic phases comprising of both $``$C3$"$ and $``$C4$"$. However, in our calculations we have considered only $``$C4$"$ for $y=0.5$ in the supercell calculations while both configurations are considered for all $y$ between $0.25$ and $0.5$ in KKR-CPA calculations. We find that for $y=0.5$, both supercell and KKR-CPA produce identical results, a high magnetic moment which is expected as the Mn spins align in $``$C4$"$. The configuration $``$C3$"$ leads to gradual quenching of the total moment as Co replaces Mn1, an extrapolation of the behaviour for $y \leq 0.25$. However, for Co replacing Ni the total moment does not change appreciably due to the proximity of the atomic moments of the two.

Thus, substitution with Co provides us with a possibility for large magnetic moment in the austenite phase when $y \sim 0.25$ and subsequently large value of $\Delta$M as desired may be realised. However, since there is a good possibility of mixed magnetic phases of the two configurations, the actual moment may not be that high as the net moment will be a weighted average of moments of the two magnetic configurations. Even then, the net magnetic moment is expected to be higher than the cases where Fe or Cu are substituted.

\subsection{Structural Phase Transition and associated change in magnetic structure} \label{MPT}

Significant MCE in the Ni-Mn based Heusler alloys is observed in the vicinity of the reversible martensitic phase transformation (MPT), a diffusionless first-order phase transition from a high symmetry austenite phase to a low symmetry martensite phase with decreasing temperature when coupled with a substantial change in the magnetic order\cite{albertini2004}. The associated large change in magnetization ($\Delta$M) across the structural phase transition  gives rise to a large magnetic entropy change i.e. an increased MCE. Thus, compositions exhibiting the structural phase transition near the room temperature and  associated with a change in magnetic structure are of great importance. Chemical substitution in Ni-Mn based ternary compounds has been proved to be an effective way to tune the stability of the austenite phase or in other words, to adjust T$_M$ and to increase $\Delta$M . As discussed in Section \ref{introduction}, different investigations conclude that Fe, Co, Cu substitutions at Mn and Ni sites in Mn-excess Ni-Mn-(Ga,In,Sn) alloys tune the thermodynamic parameters related to the magnetic and structural transformations and consequently the MCE \cite{59sahoo2011,60sahoo2014,64han2008,nayak2009,anayak2009,62feng2011}. Therefore, in this section, we have systematically investigated the effects of substitution of different elements with different concentrations and at various sites on the stability of the austenite phase of Ni$_{2}$Mn$_{1.5}$Sb$_{0.5}$. We also look into the possible changes in the magnetic configurations due to the structural transitions from cubic austenite to a tetragonal martensitic phase that can result in a large ($\Delta$M) in these compounds. Although, the quantification of $\Delta$M and comparison with experiments cannot be  directly done in this way due to the fact that the experimental samples may not lead to the tetragonal martensites immediately after the MPT and at the temperatures where experimental measurements were carried out, the calculations surely can provide important insights into the possible trends and outcomes in regard to expectation of large $\Delta$M.

\begin{figure*}[t]
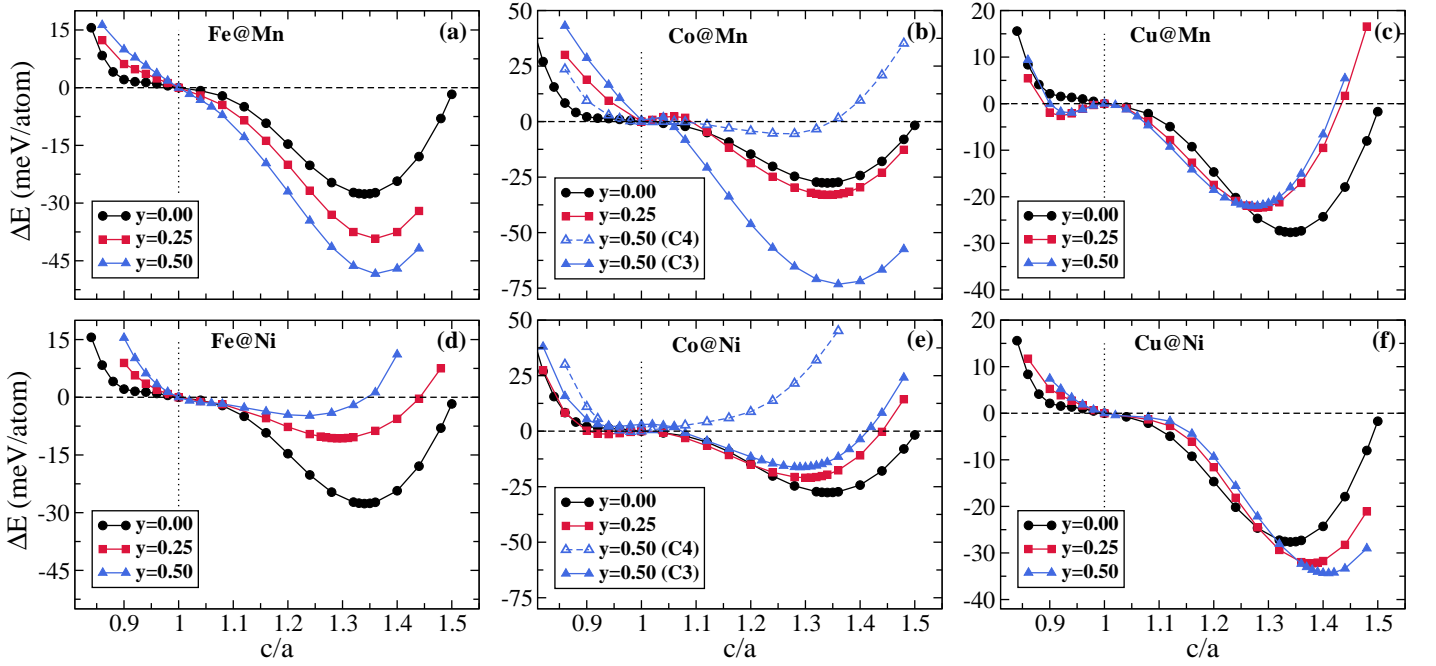

\centerline{\hfill
\psfig{file=fig2a.eps,width=0.35\textwidth}
\hspace{0.005cm}
\psfig{file=fig2b.eps,width=0.33\textwidth}
\hspace{0.005cm}
\psfig{file=fig2c.eps,width=0.33\textwidth}
\hfill}
\vspace{-0.55cm}
\centerline{\hfill
\psfig{file=fig2d.eps,width=0.35\textwidth}
\hspace{0.005cm}
\psfig{file=fig2e.eps,width=0.33\textwidth}
\hspace{0.005cm}
\psfig{file=fig2f.eps,width=0.33\textwidth}
\hfill}
\caption{The variations of total energy difference ($\Delta$E) between the austenite(L2$_{1}$) and the martensite(tetragonal) phases as a function of tetragonal distortion i.e. $c/a$ ratio for (a) Ni$_{2}$Mn$_{1.50-y}$Z$_{y}$Sb$_{0.50}$(Z@Mn) and (b) Ni$_{2-y}$Z$_{y}$Mn$_{1.50}$Sb$_{0.50}$(Z@Ni) (Z=Fe, Co and Cu) systems for considered values of $y$ and at their ground state magnetic configurations (as in Table \ref{table1}). Results where the magnetic configurations are different in the martensitic phases (for Co substitution with $y=0.50$) are also shown.} 
\label{fig2}
\end{figure*}

\begin{table*}[t]
\centering
\caption{\label{table2} Calculated values of equilibrium lattice constant (a$_{0}$), and total magnetic moment (M$_{\rm A}$) of the  systems under considerations in their austenite ground states are shown. The total energy difference ($\Delta{E}$) between the austenite(L2$_{1}$) and the martensite(tetragonal) phases $[$the equilibrium value of tetragonal distortion (c$/$a) is given in parentheses$]$, the total magnetic moment (M$_{\rm M}$) in the non-modulated martensitic phases and the differences in magnetic moments between the austenite and martensite phases ($\Delta$M) are also shown.}
\resizebox{0.86\textwidth}{!}{%
\begin{tabular}{l@{\hspace{0.4cm}} c@{\hspace{0.4cm}} c@{\hspace{0.4cm}}  c@{\hspace{0.4cm}} c@{\hspace{0.4cm}} c@{\hspace{0.4cm}}  c@{\hspace{0.4cm}} c@{\hspace{0.4cm}} }
\hline\hline
\vspace{-0.33 cm}
\\ Composition & Mag. Config. &  a$_{0}$ &  $\Delta$E($c/a$) &   M$_{\rm A}$ & M$_{\rm M}$ & $|\Delta{M}|$  \\
               &              &  (\r{A}) &  (meV/atom) &  ($\mu_{\rm B}/f.u.$)  & ($\mu_{\rm B}/f.u.$) & ($\mu_{\rm B}/f.u.$) &    \\ \hline\hline
Ni$_{2}$Mn$_{1.50}$Sb$_{0.50}$               & C1      &   5.94 &  27.64(1.34)  & 1.71 & 1.55 & 0.16   \\  
Ni$_{2}$Mn$_{1.25}$Fe$_{0.25}$Sb$_{0.50}$    & C3      &  5.92  & 39.23(1.36) &  1.51  & 1.46 & 0.05    \\
Ni$_{2}$MnFe$_{0.50}$Sb$_{0.50}$             & C3      &  5.91  & 48.38(1.36)  & 1.33  & 1.37 & 0.04    \\
\hline
Ni$_{1.75}$Fe$_{0.25}$Mn$_{1.50}$Sb$_{0.50}$ & C3      &  5.92 & 10.68(1.29)  & 2.09  & 1.88 & 0.21    \\
Ni$_{1.50}$Fe$_{0.50}$Mn$_{1.50}$Sb$_{0.50}$ & C3     &   5.91  & 4.81(1.24)  & 2.48  &   -  &   -      \\

\hline\hline

Ni$_{2}$Mn$_{1.25}$Co$_{0.25}$Sb$_{0.50}$    & C3      &  5.89 & 33.00(1.34)  & 1.04  & 0.93 & 0.11    \\
Ni$_{2}$MnCo$_{0.50}$Sb$_{0.50}$             & C4     &  5.87  & 73.26(1.36) & 4.53  & 0.09 & 4.44    \\
\hline
Ni$_{1.75}$Co$_{0.25}$Mn$_{1.50}$Sb$_{0.50}$ & C3      &  5.93  & 20.97(1.30) & 1.99  & 1.94 & 0.05     \\
Ni$_{1.50}$Co$_{0.50}$Mn$_{1.50}$Sb$_{0.50}$ & C4      &  5.95  & 16.27(1.29)  & 6.82  & 1.95 & 4.87    \\

\hline\hline

Ni$_{2}$Mn$_{1.25}$Cu$_{0.25}$Sb$_{0.50}$    & C3 &  5.92  & 22.30(1.28) & 2.94 & 3.06 & 0.12    \\
Ni$_{2}$MnCu$_{0.50}$Sb$_{0.50}$             & C4      &  5.90  & 22.00(1.27)  & 4.15 & 4.15 & 0.00    \\
\hline
Ni$_{1.75}$Cu$_{0.25}$Mn$_{1.50}$Sb$_{0.50}$ & C3  &  5.95  & 32.23(1.38)  & 1.69 & 1.64 & 0.05     \\
Ni$_{1.50}$Cu$_{0.50}$Mn$_{1.50}$Sb$_{0.50}$ & C3  &  5.98  & 34.35(1.41)  & 1.73 & 1.76 & 0.03     \\

\hline\hline
\end{tabular}
}
\end{table*}

To study the structural phase stability for a particular composition, we have distorted the lowest energy L2$_{1}$ structure at that composition along each of the possible crystallographic inequivalent directions and computed the total energy as a function of the tetragonal distortion given by $(c/a)$. Due to finite size of the 16 atom supercell, different atomic distributions in the planes, perpendicular to which distortion is given, define two crystallographic inequivalent directions for most of the compositions considered here. Further, for the purpose of investigation of any possible changes in the magnetic structures in the martensitic phases from that in the cubic phases, we have also calculated energy profiles for the possible magnetic configurations listed in Table \ref{table1} at each composition. The results are shown in Figure 1 of supplementary material. In each of these plots, the reference energy is the energy of the austenite ground state for the corresponding composition. In Figure \ref{fig2} we summarise the results and show the energy profiles of the configurations which provide the minimum energy in each case.

Figure \ref{fig2}(a) shows that, when Fe is substituted at Mn site in Ni$_{2}$Mn$_{1.5}$Sb$_{0.5}$, the stability of the austenite phase decreases with increasing concentration of Fe as is apparent from the growing difference in energy ($\Delta$E) (Table \ref{table2}) between the austenite and the martensite phases, implying that T$_{M}$ increases with Fe concentration. This is completely in contrast with the qualitative behaviour observed in the experiment \cite{59sahoo2011}. We address this anomaly in detail as a special case in Section \ref{Fe-doped}. Fe substituting Ni, on the other hand, enhances the stability of the austenite phase. For 50\% Fe-substitution, the total energy curve has a shallow minima and thus the possibility of a martensitic instability is slim. Thus the relative phase stability due to Fe substitution is dependent on the atom that is being substituted. For both cases, no changes in magnetic structures have been observed near the structural phase transition and as a result $\Delta$M are also found to be very small. Substitution of Cu, in place of either Ni or Mn does not affect the relative phase stability significantly. As the figures \ref{fig2}c and \ref{fig2}f suggest, $\Delta$E do not change appreciably with $y$, implying that T$_{M}$ remains almost unchanged even when concentration of the substituents are high. In both cases, the magnetic structures do not change across phases as is clear from the results in Table \ref{table2}. Co-substitution turns out to be very interesting as compared to the other two cases.
With increase of Co concentration substituting Mn, $\Delta$E first increases moderately in the lower concentration range ($y \leq 0.25$) and no changes is magnetic structure across phases are observed resulting in a lower value of $\Delta$M. Further increase in Co concentration brings in a change in the magnetic structure in austenite phase as was discussed in the previous section; however, the magnetic structure remains same in the martensitic phase with Mn spins still aligned anti-parallel, leading to a $\Delta$M value of 4.44~$\mu_{\rm B}/f.u.$. Simultaneously, we find a substantial increase in $\Delta$E implying a greater martensitic instability in the system. Large $\Delta$M coupled with a large T$_{M}$ indicate better prospect for MCE. Similar jump in $\Delta$M is observed in case of Co substituting Ni. However, the $\Delta$E decreases in this case with $y$. This is consistent with the experimental observation in Ni$_{2-y}$Co$_{y}$Mn$_{1.52}$Sb$_{0.48}$ with $y=0.2$ \cite{nayak2009}. Incidentally, large MCE has been found in this compound. Thus, our results in Co-substituted compounds indicate that they are potential materials for large MCE near a magneto-structural transition of first order.

\subsection{Magnetic exchange interactions} \label{jij}

\begin{figure*}[t]
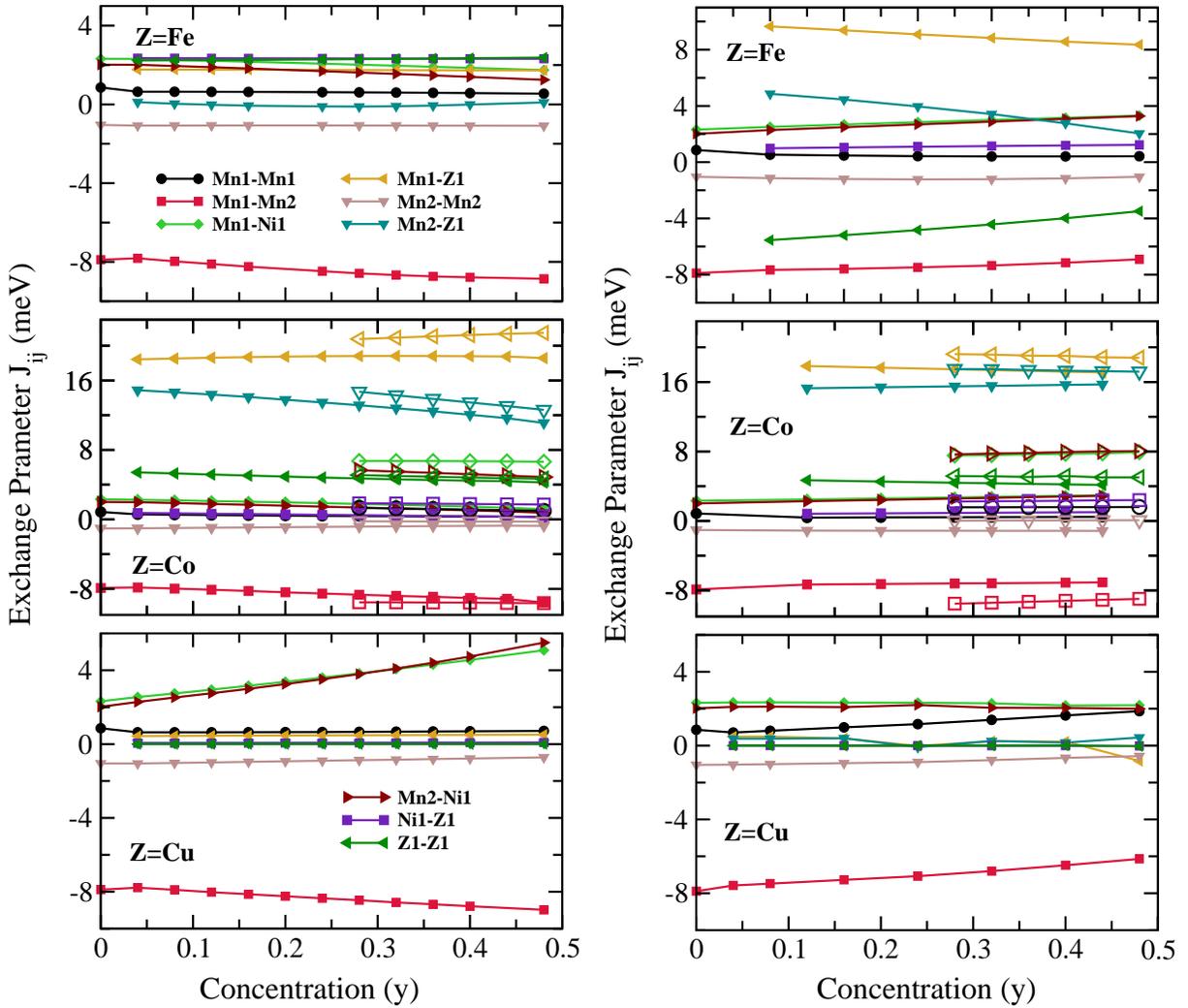

\centerline{\hfill
\psfig{file=fig3a.eps,width=0.44\textwidth}
\hspace{0.05cm}
\psfig{file=fig3b.eps,width=0.44\textwidth}
\hfill}
\caption{The dependence of the inter-atomic magnetic exchange parameters in the first coordination shell (for different pair of atoms) for (left panel) Ni$_{2}$Mn$_{1.52-y}$Z$_{y}$Sb$_{0.48}$(Z@Mn) and (right panel) Ni$_{2-y}$Z$_{y}$Mn$_{1.52}$Sb$_{0.48}$(Z@Ni) with Z=Fe, Co and Cu systems in their austenite phases. Calculations in each case, are done at the ground state magnetic configuration. The composition of the parent compound ($y=0$), in each case, is considered to be the ones in the experiments \cite{nayak2009,60sahoo2014,64han2008,anayak2009,59sahoo2011,62feng2011}, and are nominally different from that of Ni$_{2}$Mn$_{1.5}$Sb$_{0.5}$. For Co-substituted systems, magnetic exchange parameters both for $``$C3$"$(closed symbols) and $``$C4$"$(open symbols) magnetic configurations for $y>0.25$ are shown.}
\label{fig3}
\end{figure*}

The inter-atomic magnetic exchange interactions provide the understanding behind the evolution of magnetic transitions and the occurrence of MCE in cases of magneto-structural transitions as observed in Ni-Mn-Z compounds \cite{buchelnikov2008,singh2013,55sokolovskiy2015,sokolovskiy2013,sokolovskiy2014,moya2007,79krenke2005,22khan2007}. In this section, we investigate the effects of substitutions of different constituents of Ni$_{2}$Mn$_{1.5}$Sb$_{0.5}$ at different sites and with different concentrations, on the magnetic exchange interactions, in their austenite phases, and try to correlate with the results obtained in Sections \ref{moment} and \ref{MPT}. We then exclusively discuss the reasons behind large MCE observed in experiment on Ni$_{2-y}$Co$_{y}$Mn$_{1.52}$Sb$_{0.48}$ in light of these. 

Figure \ref{fig3} shows the trends in variations of various nearest neighbour inter-atomic magnetic exchange coupling strengths as a function of the substituents concentration. Only the variations in the first coordination shells are considered as these are the dominant interactions. We find that, in each case of substitution, the overall ferromagnetic interactions increase due to the predominantly ferromagnetic interactions between the Ni-Z, Mn-Z and Z-Z pairs. When Mn is substituted by Fe there is a competition between the ferromagnetic coupling of Ni-Fe, Mn1-Fe, and antiferromagnetic coupling of Mn1-Mn2 pairs. The antiferromagnetic interaction of Mn1-Mn2 pairs increases with concentration of Fe and compensates the weak increase in the ferromagnetic interactions with $y$. The small negative changes in the magnetic moment with $y$ can be correlated to such variations in the exchange interactions. In a complete contrast to this, substitution of Fe at Ni sites amplify the ferromagnetic Mn-Fe interactions, along with a simultaneous weakening of the antiferromagnetic interactions(Mn1-Mn2, Fe-Fe) as $y$ increases. The increase in the magnetic moment with $y$ is an artefact of this. Almost no variations in the magnetic moment of compounds when Ni is substituted by Cu can be understood from the minimal variations in both ferromagnetic and antiferromagnetic exchange interactions with Cu concentration. In contrast, The significant strengthening in Ni-Mn ferromagnetic interactions when Cu substitutes Mn, can be correlated to the increase in the magnetic moment of the corresponding compound.
Co substitutions, both at Mn and Ni sites, give rise to the largest ferromagnetic coupling strengths which is due to very strong ferromagnetic exchange interactions between the Co and Mn atoms. ferromagnetic interactions between Mn-Ni and Co-Co pairs strengthen it further. For higher concentrations ($y > 0.25$) the parallel alignment of the Mn atoms (as in $``$C4$"$ magnetic configuration) magnifies the ferromagnetic interactions further. This explains the large value of moment at high concentrations of Co. Overall, it can be concluded that substitution of magnetic 3d-elements in Ni$_{2}$Mn$_{1.5}$Sb$_{0.5}$ magnify ferromagnetic exchange interactions in the system, and thus in general, leads to a higher value of magnetic moment with respect to the parent compound. 
\begin{figure*}[t]
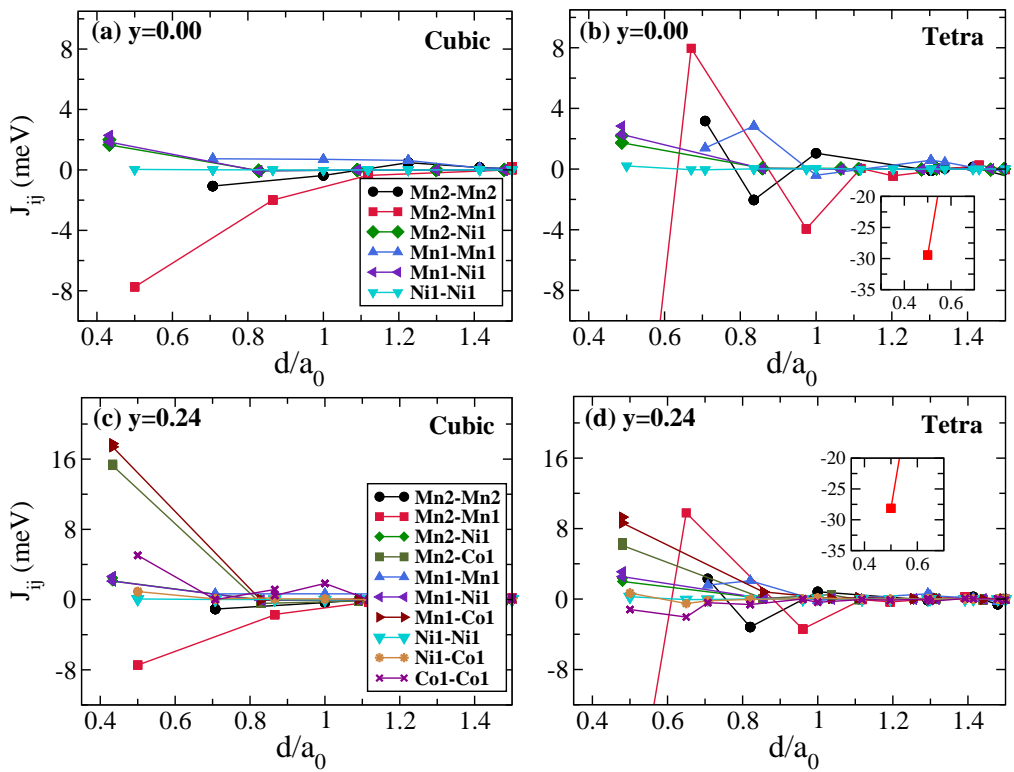

\centerline{\hfill
\psfig{file=fig4a.eps,width=0.378\textwidth}
\hspace{0.05cm}
\psfig{file=fig4b.eps,width=0.35\textwidth}
\hfill}
\vspace{0.05cm}
\centerline{\hfill
\psfig{file=fig4c.eps,width=0.378\textwidth}
\hspace{0.05cm}
\psfig{file=fig4d.eps,width=0.35\textwidth}
\hfill}
\caption{Inter-atomic magnetic exchange interactions in (a)-(b)parent composition Ni$_{2}$Mn$_{1.52}$Sb$_{0.48}$ and (c)-(d) Ni$_{2-y}$Co$_{y}$Mn$_{1.52}$Sb$_{0.48}$ with $y=0.24$ in
the cubic ($c/a=1$) and tetragonal ($c/a \neq 1$) phases as a function of distance d(in units of lattice constant a$_{0}$) between the pair of atoms.}
\label{fig4}
\end{figure*}

We next focus on understanding of the occurrence of large $\Delta$M and the possible connection to significant MCE in some of the substituted Mn-excess, Sb-deficient Ni$_{2}$MnSb compounds. For this purpose, we consider Ni$_{2}$Mn$_{1.52}$Sb$_{0.48}$ compound and investigate the behaviour of magnetic exchange interactions, in both austenite and martensite phases, when Fe, Co and Cu substitute Ni and Mn with the concentration of the substituent being $0.24$. The reason for picking this particular composition is the observation of large MCE for compositions close to this \cite{anayak2009} in the compound where Co substitutes Ni. In Figure \ref{fig4}, we show the results for Ni$_{1.76}$Co$_{0.24}$Mn$_{1.52}$Sb$_{0.48}$ only. Results for other compounds are shown in Figures 2-5 of supplementary materials. In the parent composition Ni$_{2}$Mn$_{1.52}$Sb$_{0.48}$, the austenite phase is dominated by the Mn1-Mn2 nearest-neighbour antiferromagnetic (AFM) interactions. In the tetragonal martensite phase this AFM interaction gets almost four times magnified and shows an oscillatory behaviour in the second and third coordination shells. As a result not much difference between the magnetic moments in the two phases are expected to occur. This proposition is consistent with the results on Ni$_{2}$Mn$_{1.5}$Sb$_{0.5}$ (Table \ref{table2}). When Co substitutes Ni, the Mn1-Mn2 AFM interaction strength remains same as the pristine compound in the martensitic phase while the ferromagnetic interactions gain slightly. But, in the austenite phase, the Co-Mn ferromagnetic interactions amplify more than five times in comparison to the dominant ferromagnetic interactions in the pristine compound. As a result the magnetic moments in the two phases would differ substantially, in comparison to that in the pristine compound. This, therefore, perfectly explains the experimentally observed large $\Delta$M and large MCE. Another highlight of these calculations is observation of a correlation between the qualitative nature of the variations in the exchange interactions and relative stabilities of the structural phases. Experimentally it was observed that  for Co-substitution at Ni site in Ni$_{2}$Mn$_{1.52}$Sb$_{0.48}$,T$_{M}$ decreases with the substituent concentration implying that this substitution stabilises the austenite phase. The nature of magnetic exchange interactions in Figure \ref{fig4} suggests that the strong ferromagnetic interactions stabilise the austenite phase. The analysis of Figures 2-5 in supplementary materials corroborate this.

\subsection{Mechanical properties} \label{elastic}
 MCE materials with coupled magneto-structural transitions often suffer from cracking and fatigue, which severely limits their usefulness. Quaternary off-stoichiometric Ni-Mn based Heusler compounds are found to exhibit better mechanical properties such as ductility, corrosion resistance, machinability, all of which ease manufacturing and increase the product longevity. For example, Fe-addition improves the toughness of Ni-Mn-Ga allloys without sacrificing its magnetic and thermoelastic properties\cite{cherechukin2001,28bkundu2017}. In this section, we, therefore, explore the changes in the mechanical properties of Ni$_{2}$Mn$_{1.5}$Sb$_{0.5}$ compounds upon substitution by the fourth element. \\

\begin{table}[t]
\centering
\caption{\label{table3} The calculated values of Pugh ratio G$_v$/B and Cauchy pressure C$^P$ of the systems under considerations in their austenite phases with ground state magnetic configurations for corresponding compositions.}
\resizebox{0.5\textwidth}{!}{%
\begin{tabular}{l@{\hspace{0.4cm}} c@{\hspace{0.4cm}} l@{\hspace{0.4cm}} c@{\hspace{0.4cm}} }
\hline\hline
\vspace{-0.33 cm}
\\ Composition & Mag. Config. &  G$_v$/B  & C$^P$(GPa)   \\
\hline
Ni$_{2}$Mn$_{1.50}$Sb$_{0.50}$               & C1       &  0.39  &  49.19      \\  
Ni$_{2}$Mn$_{1.25}$Fe$_{0.25}$Sb$_{0.50}$    & C3       &  0.43  &  41.77     \\
Ni$_{2}$MnFe$_{0.50}$Sb$_{0.50}$             & C3       &  0.44  &  38.93     \\
\hline
Ni$_{2}$Mn$_{1.50}$Sb$_{0.50}$               & C1       &  0.39  &  49.19    \\
Ni$_{1.75}$Fe$_{0.25}$Mn$_{1.50}$Sb$_{0.50}$ & C3       &  0.47  &  31.68    \\
Ni$_{1.50}$Fe$_{0.50}$Mn$_{1.50}$Sb$_{0.50}$ & C3       &  0.49  &  25.31   \\

\hline\hline

Ni$_{2}$Mn$_{1.50}$Sb$_{0.50}$               & C1       &  0.39  &  49.19  \\
Ni$_{2}$Mn$_{1.25}$Co$_{0.25}$Sb$_{0.50}$    & C3       &  0.42  &  42.70   \\
Ni$_{2}$MnCo$_{0.50}$Sb$_{0.50}$             & C4       &  0.45  &  40.12  \\
\hline
Ni$_{2}$Mn$_{1.50}$Sb$_{0.50}$               & C1       &  0.39  & 49.19  \\
Ni$_{1.75}$Co$_{0.25}$Mn$_{1.50}$Sb$_{0.50}$ & C3       &  0.43  & 40.85   \\
Ni$_{1.50}$Co$_{0.50}$Mn$_{1.50}$Sb$_{0.50}$ & C4       &  0.41  & 44.81  \\

\hline\hline

Ni$_{2}$Mn$_{1.50}$Sb$_{0.50}$               & C1       &  0.39  & 49.19  \\
Ni$_{2}$Mn$_{1.25}$Cu$_{0.25}$Sb$_{0.50}$    & C3       &  0.46  & 33.78  \\
Ni$_{2}$MnCu$_{0.50}$Sb$_{0.50}$             & C4       &  0.39  & 52.65   \\
\hline
Ni$_{2}$Mn$_{1.50}$Sb$_{0.50}$               & C1       &  0.39  & 49.19   \\
Ni$_{1.75}$Cu$_{0.25}$Mn$_{1.50}$Sb$_{0.50}$ & C3       &  0.42  & 41.86   \\
Ni$_{1.50}$Cu$_{0.50}$Mn$_{1.50}$Sb$_{0.50}$ & C3       &  0.41  & 42.05   \\

\hline\hline
\end{tabular}
}
\end{table}

A good measure of whether the system is more ductile or more brittle is its Pugh ratio given as G$_v$/B\cite{92pugh1954,50roy2015}, Gv is the isotropic shear modulus under Voigt formalism\cite{49voigt1889} related to the resistance of the material to plastic deformation and B is the bulk modulus in the cubic phase. Compounds having a Pugh ratio greater than 0.57 are considered to be more brittle. On the other hand, Cauchy pressure C$^P$, defined as C$^p$=(C$_{12}$-C$_{44}$), provides insight to the nature of bonding in a material with cubic symmetry\cite{94pettifor1992}; C$_{12}$, C$_{44}$ are the shear moduli in the cubic phase. A positive value of Cauchy pressure indicates the bonding in the compound to be more metallic while a negative value implies a stronger covalent bonding\cite{28bkundu2017,ghosh2019}. To calculate the Pugh ratio and Cauchy pressure, at first we have calculated the Bulk modulus(B) and the elastic moduli (C$_{44}$ and C$^{\prime}$) for all the compounds considered in this work, in their austenite phases (the results are shown in Figure 6 and in TableI of supplementary material). We find that the bulk modulus decreases for all the cases where Mn is substituted. The bulk modulus does not change when Fe and Co substitute Ni but decreases appreciably when Cu replaces Ni. These trends in B are consistent with the trends in the variations of lattice constants as shown in Table \ref{table2}. Positive C$_{44}$ for all the compositions satisfies one of the stability criteria for cubic crystals. A negative C$^{\prime}$ indicates the instability in the L2$_{1}$ phase and that the system is prone to a structural transformation. The values of C$^{\prime}$ tabulated in TableI of supplementary material explains the observed martensitic instabilities in the compounds and will be discussed later in Section \ref{Tm-Tc} in detail. From the calculated bulk modulus and C$^{\prime}$, we have calculated C$_{11}$ and C$_{12}$,  and then G$_v$/B and C$^P$ using the relations in Section \ref{methods} for all the compounds. The results are shown in Table \ref{table3}. The results imply that substitution of Fe, Co or Cu keeps the Ni-Mn-Sb compounds ductile and the nature of bonding largely metallic.

\subsection{Variation in T$_{M}$ and T$_c^A$} \label{Tm-Tc}
A large MCE i.e. $\Delta$S$_{M}$ is usually obtained at a temperature near T$_{M}$ in cases of first-order magneto-structural transitions, typical for Ni-Mn based Heusler compounds. However small $\Delta$S$_{M}$ may also be observed near T$_c^A$, the magnetic transition temperature in the austenite phase, in a second order magnetic transition. The largest MCE can be obtained if these two temperatures are as close as possible and near room temperature, for operational purpose. Attempts\cite{vasil1999,aliev2004,albertini2004}, thus, have been made to bring these two temperatures closer by adjusting the composition so that T$_{M}$ increases and  T$_c^A$ decreases. Success in this approach has been achieved in case of Ni$_{2.18}$Mn$_{0.82}$Ga by substitution of the magnetic components with another transition metal from the $3d$ series \cite{cherechukin2004,khan2005,soto2008,stadler2006,liu2002}. After investigating the roles of the substituents in achieving a large $\Delta$M and the underlying physics therein in the previous sub-section, thus pinpointing the materials which can potentially be exhibiting significant MCE, it becomes necessary to explore how T$_{M}$ and T$_c^A$ behave with substitution of Fe, Co or Cu in Ni$_{2}$Mn$_{1.5}$Sb$_{0.5}$; more so as the experimental results on Ni-Mn-Sb compounds with compositions close to the parent compound considered in this work suggest that the two temperatures are quite close \cite{62feng2011,64han2008,anayak2009,60sahoo2014} and near room temperature, varying between 260-330K.

\begin{table*}[t]
\centering
\caption{\label{table4} Calculated values all predicting observables of martensitic transition temperature, T$_{M}$: electron to atom ratio ($e/a$),  electron density (n), total energy difference ($\Delta{E}$) between the austenite(L2$_{1}$) and the martensite(tetragonal) phases, tetragonality of the martensite phase ($\mid c/a-1\mid$) and shear modulus (C$^\prime$) of the austenite phase for all the six types of considered systems. In the last column trend in T$_{M}$ and Curie temperature (T$_c^A$) in the austenite phase have been concluded by observing the trend in more reliable quantity C$^\prime$ and values of T$_c^A$ calculated through Monte Carlo Simulation in Figure \ref{fig5}, respectively.}
\resizebox{0.93\textwidth}{!}{%
\begin{tabular}{l@{\hspace{0.4cm}} c@{\hspace{0.4cm}} l@{\hspace{0.4cm}} c@{\hspace{0.4cm}} c@{\hspace{0.4cm}} c@{\hspace{0.4cm}} c@{\hspace{0.4cm}} c@{\hspace{0.4cm}} }
\hline\hline
\vspace{-0.33 cm}
\\ Composition & Mag. Config. & $e/a$ & n & $\Delta$E  &  $\mid c/a-1\mid$ & C$^\prime$ & Trends in T$_{M}$ and T$_c^A$ \\
               &              &       &   & (meV/atom) &                   & (GPa)      &                               \\
\hline
Ni$_{2}$Mn$_{1.50}$Sb$_{0.50}$               & C1       &  8.25    &  0.630  &  27.64 & 0.34 & -3.81       & T$_{M}$ decreases,      \\  
Ni$_{2}$Mn$_{1.25}$Fe$_{0.25}$Sb$_{0.50}$    & C3       &  8.3125  &  0.641  &  39.23 & 0.36 &  1.65       & T$_c^A$ decreases slightly \\
Ni$_{2}$MnFe$_{0.50}$Sb$_{0.50}$             & C3       &  8.375   &  0.649  &  48.38 & 0.36 &  5.04       &  \\
\hline
Ni$_{2}$Mn$_{1.50}$Sb$_{0.50}$               & C1       &  8.25    &  0.630  &  27.64 & 0.34 & -3.81       & T$_{M}$ decreases, \\
Ni$_{1.75}$Fe$_{0.25}$Mn$_{1.50}$Sb$_{0.50}$ & C3       &  8.125   &  0.627  &  10.68 & 0.29 &  5.28       & T$_c^A$ decreases \\
Ni$_{1.50}$Fe$_{0.50}$Mn$_{1.50}$Sb$_{0.50}$ & C3       &  8.00    &  0.620  &  4.81  & 0.24 &  13.12      & \\

\hline\hline

Ni$_{2}$Mn$_{1.50}$Sb$_{0.50}$               & C1       &  8.25    &  0.630  &  27.64 & 0.34 & -3.81        & T$_{M}$ increases,  \\
Ni$_{2}$Mn$_{1.25}$Co$_{0.25}$Sb$_{0.50}$    & C3       &  8.375   &  0.656  &  33.00 & 0.34 & -9.26        & T$_c^A$ decreases 
\\
Ni$_{2}$MnCo$_{0.50}$Sb$_{0.50}$             & C4       &  8.50    &  0.672  &  73.26 & 0.36 & -14.06 &\\
\hline
Ni$_{2}$Mn$_{1.50}$Sb$_{0.50}$               & C1       &  8.25    &  0.630  &  27.64 & 0.34 & -3.81        & T$_{M}$ decreases, 
\\
Ni$_{1.75}$Co$_{0.25}$Mn$_{1.50}$Sb$_{0.50}$ & C3       &  8.1875  &  0.628  &  20.97 & 0.30 & -0.55        & T$_c^A$ increases slightly  \\
Ni$_{1.50}$Co$_{0.50}$Mn$_{1.50}$Sb$_{0.50}$ & C4       &  8.125   &  0.617  &  16.27 & 0.29 & -1.90        & \\

\hline\hline

Ni$_{2}$Mn$_{1.50}$Sb$_{0.50}$               & C1       &  8.25    & 0.630   &  27.64 & 0.34 & -3.81        & T$_{M}$  decreases in the beginning \\
Ni$_{2}$Mn$_{1.25}$Cu$_{0.25}$Sb$_{0.50}$    & C3       &  8.50    & 0.656   &  22.30 & 0.28 & 15.64        & and then increases, \\
Ni$_{2}$MnCu$_{0.50}$Sb$_{0.50}$             & C4       &  8.75    & 0.682   &  22.00 & 0.27 & -7.95        & T$_c^A$ decreases \\
\hline
Ni$_{2}$Mn$_{1.50}$Sb$_{0.50}$               & C1       &  8.25    & 0.630   &  27.64 & 0.34 & -3.81        & T$_{M}$ and T$_c^A$ remains \\
Ni$_{1.75}$Cu$_{0.25}$Mn$_{1.50}$Sb$_{0.50}$ & C3       &  8.3125  & 0.631   &  32.23 & 0.38 & -1.51        & almost constant \\
Ni$_{1.50}$Cu$_{0.50}$Mn$_{1.50}$Sb$_{0.50}$ & C3       &  8.375   & 0.627   &  34.35 & 0.41 & -1.26        & \\

\hline\hline
\end{tabular}
}
\end{table*}

\begin{figure}[t]
\centerline{\hfill
\psfig{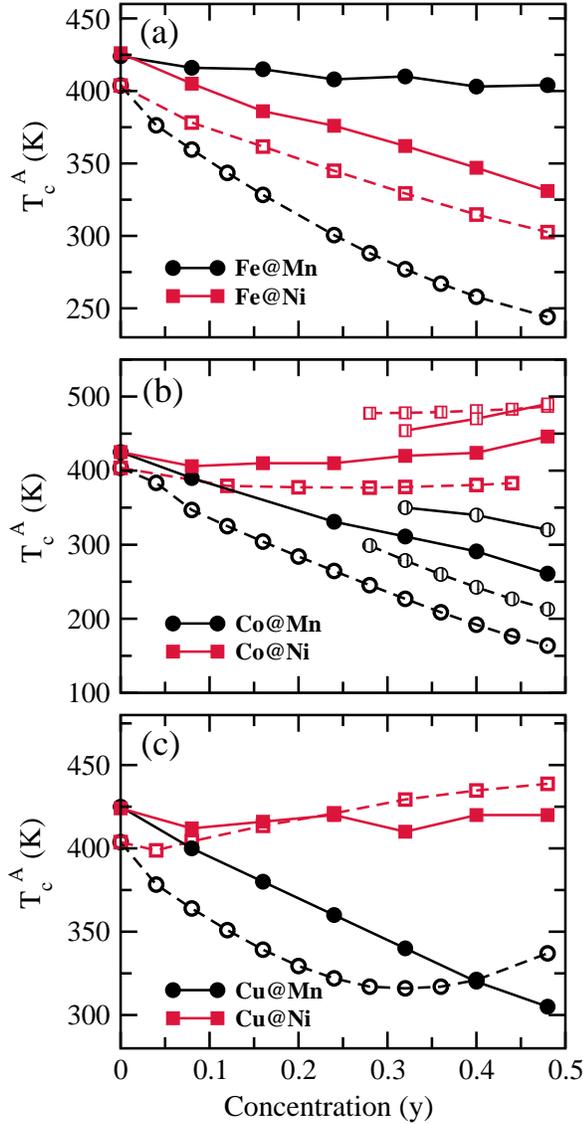}
\hfill}
\caption{Calculated Curie temperatures (T$_c^A$) as a function of substituents (Z=Fe, Co, Cu) concentration for Ni$_{2}$Mn$_{1.52-y}$Z$_{y}$Sb$_{0.48}$(Z@Mn) and Ni$_{2-y}$Z$_{y}$Mn$_{1.52}$Sb$_{0.48}$(Z@Ni) systems. Closed symbols and Open symbols represent results calculated by Monte Carlo Simulation (MCS) and Mean Field Approximation (MFA) methods, respectively. For Co-substituted systems, Curie temperatures in $``$C4$"$ magnetic configuration for $y>0.25$ are shown with marked circles.}
\label{fig5}
\end{figure}

Calculation of T$_{M}$ requires computations of  free energies including contributions from phonon and magnetic excitations, apart from the electronic one. This becomes computationally prohibitive for off-stoichiometric compounds like the ones considered here. Therefore, in this work, we could only look at the qualitative variations in T$_{M}$ as a function of composition for the compounds considered here, by studying the variations of quantities which are related to T$_{M}$\cite{li2011,25hu2009,45li2011,68siewert2012}, and are calculated here. The $e/a$ ratio has been found to be a good predictor for composition dependence of T$_{M}$ in Ni-Mn based Heusler alloys. T$_{M}$ is found to follow $e/a$ \cite{29lanska2004,30zayak2006,31mehaddene2008}, albeit with exceptions in some cases: Fe substituted Ni-Mn-Sb being one\cite{59sahoo2011}. For a few notable cases like N-Mn-Ga compounds \cite{45li2011}, the electron density $n=((e/a)\times$n$_{1}$)/V$_{cell}$, scales with T$_{M}$, n$_{1}$ being the average number of atoms contained in the unit cell of volume V$_{cell}$\cite{chen2007}. The total energy difference($\Delta$E) between the austenite and the martensite phases is found to be another predictor of T$_{M}$\cite{89chakrabarti2013,81sokolovskiy2017,28bkundu2017,ghosh2019}. Larger value of $\Delta$E implies a higher stability of the martensite phase at zero temperature and thus a higher T$_{M}$. The tetragonality of the martensite phase quantified by $\mid c/a-1\mid$ is also found to follow variations in T$_{M}$ in cases of Ni-Mn-(Ga,In) compounds \cite{29lanska2004,banik2007,27li2012}. The best predictor of T$_{M}$ for Ni-Mn based Heusler alloys, so far, has been the shear modulus C$^\prime$ in the austenite phase. This is due to the fact that the shear modulus is associated with the softening of the acoustic phonon branch that drives the martensitic transformation in these systems. Important physical factors like site ordering and magnetic structure are all taken care of in the variation of this quantity \cite{36ren2000,37ren2003}.

In Table \ref{table4}, we compile the trends in variations of these five quantities as a function of composition for the six compounds considered, to qualitatively understand the variations in T$_{M}$. With increase in Fe concentration at the expense of Mn, the $e/a$ ratio, electron density n, $\mid c/a-1\mid$ and $\Delta$E increases, suggesting increase in T$_{M}$ with Fe concentration, while the increase in C$^{\prime}$ with concentration of Fe suggests the stabilisation of the austenite phase, an opposite trend. This later trend is in agreement with the experimental observation \cite{59sahoo2011} which shows a decrease in T$_{M}$ with Fe concentration. Even though we consider the trends in C$^{\prime}$ as the authentic one, this particular system exhibits discrepancy between theory and experiment with regard to the trend in the magnetic moment in the austenite phase, as mentioned earlier. We discuss the origin of this discrepancy and possible solution in Section \ref{Fe-doped}. For compounds with Fe substituting Ni, the trends in all five quantities are consistent, indicating lowering of T$_{M}$ with Fe concentration. Though there is no experimental observations available on compositions close to Ni$_{2-y}$Fe$_{y}$Mn$_{1.5}$Sb$_{0.5}$, Fe substitution at Ni sites for similar systems  Ni-Mn-(Ga, In, Sn) show the same trend in variation of T$_{M}$\cite{krenke2007,soto2008}. Our result for this system, therefore, is consistent. In case of Co-substitution, irrespective of whether Mn or Ni is substituted, all five quantities show the same trend which implies that T$_{M}$ increases(decreases) with Co concentration, when Co substitutes Mn(Ni). The available experimental results \cite{nayak2009,anayak2009,64han2008} for Ni$_{2-y}$Co$_{y}$Mn$_{1.52}$Sb$_{0.48}$ are in agreement with this. For Cu substitution at Mn site, both $e/a$ and n increase wheres $\Delta$E decreases slightly, implying contrasting trends in T$_{M}$. On the other hand, C$^{\prime}$ initially increases with Cu for the low concentrations, consistent with the trend seen in $\Delta$E, predicting a slight decrease in T$_{M}$, only to decrease for higher concentrations. For compounds where Cu substitutes Ni, $\Delta$E and C$^\prime$, too, show opposite trends. However changes in all the quantities are very small indicating that the T$_{M}$ remains almost constant with respect to the parent compound. In absence of experiments on this system, this cannot be verified. Nevertheless, it is noteworthy that other Ni-Mn based Heulser compounds show almost the same trend for Cu substitution\cite{ezekiel2018,khan2005}.     \\

The variation in Curie temperature (T$_c^A$) in the austenite phase for all the considered systems, calculated by Mean field approximation (MFA) and Monte Carlo simulation (MCS) methods, have been shown in Figure \ref{fig5}. From Figure \ref{fig5}, it can be seen that the overall trend remains almost same for most of the cases, irrespective of the method used, and that the calculated values, in general, are overestimated in comparison to experiments. For Fe substitution at Mn site, T$_c^A$, calculated by MCS method slightly decreases with Fe concentration wheres with MFA calculation a substantial decrease in T$_c^A$ is observed. On the other hand, when Fe is substituted at Ni site, same trend of T$_c^A$ decreasing linearly with Fe concentration, is found from calculations by either method. We find the same for Co-substituted compounds. For compounds with Co substituting Mn, T$_c^A$ decreases with Co concentration when $``$C3$"$ magnetic configuration is considered. For $``$C4$"$ magnetic configuration in the higher concentration range, same trend is observed ;  the values of T$_c^A$ are larger though. For compounds with Co substituting Ni, the T$_c^A$ slightly increases with Co concentration. For compounds with Cu substitution,  T$_c^A$ decreases linearly when substitution is done at Mn site, whereas for substitution at Ni site T$_c^A$ remains almost constant when calculated with MCS method.  T$_c^A$ calculated by MFA shows a different trend for substitution at Mn site for which T$_c^A$ first decreases in the low concentration range and then it increases in the higher concentration range.

In Table \ref{table4} we have summarised the above discussion with regard to variations in T$_{M}$ and T$_{c}^{A}$. Among the compounds which showed promises as magnetocaloric materials by means of large $\Delta$M, the Co-substituted ones at concentrations of Co more than $25 \%$, Ni$_{2}$Mn$_{1.5-y}$Co$_{y}$Sb$_{0.5}$ can have T$_{M}$ and T$_{c}^{A}$ very close, desirable for large MCE. In this case, if we take into account the fact that our calculated T$_{c}^{A}$ is overestimated by around 50~K, the T$_{c}^{A}$ can be very close to room temperature for $y \sim 0.25$. On the other hand, a crude estimation of T$_{M}$ can be done from the values of $\Delta$E. If the $\Delta$E of the parent compound is mapped to 260~K, the possible T$_{M}$ by extrapolation from experimental results\cite{21j2007,22khan2007,20w2009} on compounds with compositions close to it, then, T$_{M}$ will be close to 300~K for $y \sim 0.25$. This together with large $\Delta$M will make this compound with this composition a desirable material for MCE. Ni$_{2-y}$Co$_{y}$Mn$_{1.5}$Sb$_{0.5}$ will not be as effective since its T$_{c}^{A}$ will be around 350~K for $y \sim 0.25$ while the T$_{M}$ will be around 220~K, as per the crude estimation from $\Delta$E. Although the $\Delta$M is very low, the compound Ni$_{2}$Mn$_{1.5-y}$Cu$_{y}$Sb$_{0.5}$ exhibits the possibility of almost coincidence of T$_{M}$ and T$_{c}^{A}$ as the former remains almost constant and the later rapidly decreases towards room temperature.

\subsection{Resolving discrepancy between theory and experiment for Ni$_{2}$Mn$_{1.52-y}$Fe$_{y}$Sb$_{0.48}$: possible role of site occupancy } \label{Fe-doped}

\begin{table*}[t]
\centering
\caption{\label{table5} All possible site-occupation configurations ($``$S-a$"$ to $``$S-e$"$) and corresponding ground state magnetic configurations of Ni$_{2}$Mn$_{1.50-y}$Fe$_{y}$Sb$_{0.50}$ (y=0 and 0.25) with their relative electronic energies $\Delta$E$_{0}$ (in meV/atom), considering the electronic energy of the $``$S-a$"$ configuration as reference one.}
\resizebox{1.00\textwidth}{!}{%
\vspace{0.3 cm}
\begin{tabular}{l@{\hspace{0.05cm}}  c@{\hspace{0.5cm}} l@{\hspace{0.2cm}} l@{\hspace{0.2cm}} l@{\hspace{0.2cm}} l@{\hspace{0.4cm}} c@{\hspace{0.2cm}}}
\hline\hline
\vspace{-0.33 cm}
\\ Composition & Site     & \multicolumn{3}{c}{Site Occupancies} & Mag. Configurations & $\Delta$E$_{0}$ \\
               & Config.  & 4a   &   4b   &    8c             &                             &            \\
\hline\hline
Ni$_{2}$Mn$_{1.50}$Sb$_{0.50}$ & S-a & Sb1$_{0.50}$Mn2$_{0.50}$ & Mn1 & Ni1$_{2}$ & C1 (Mn1$\uparrow$ Mn2$\downarrow$ Ni1$\uparrow$) & 0.00  \\
\hline
Ni$_{2}$Mn$_{1.25}$Fe$_{0.25}$Sb$_{0.50}$ & S-a & Sb1$_{0.50}$Mn2$_{0.25}$Fe1$_{0.25}$ & Mn1 & Ni1$_{2}$ & C3(Mn1$\uparrow$ Mn2$\downarrow$ Ni1$\uparrow$ Fe1$\uparrow$) & 0.00 \\
                                                  & S-b & Sb1$_{0.50}$Mn2$_{0.50}$ & Mn1$_{0.75}$Fe1$_{0.25}$ & Ni1$_{2}$ & C3(Mn1$\uparrow$ Mn2$\downarrow$ Ni1$\uparrow$ Fe1$\uparrow$) & -4.06 \\
                                                  & S-c & Sb1$_{0.50}$Mn2$_{0.25}$Ni2$_{0.25}$ & Mn1 & Ni1$_{1.75}$Fe1$_{0.25}$ & C3(Mn1$\uparrow$ Mn2$\downarrow$ Ni1,Ni2$\uparrow$ Fe1$\uparrow$) & 5.80 \\
                                                  & S-d & Sb1$_{0.50}$Mn2$_{0.50}$ & Mn1$_{0.75}$Ni2$_{0.25}$ & Ni1$_{1.75}$Fe1$_{0.25}$ & C3(Mn1$\uparrow$ Mn2$\downarrow$ Ni1,Ni2$\uparrow$ Fe1$\uparrow$) & -3.09 \\
                                                  & S-e & Sb1$_{0.50}$Mn2$_{0.25}$Ni2$_{0.25}$ & Mn1$_{0.75}$Fe1$_{0.25}$ & Ni1$_{1.75}$Mn3$_{0.25}$ & C3(Mn1$\uparrow$ Mn2$\downarrow$ Mn3$\uparrow$ Ni1,Ni2$\uparrow$ Fe1$\uparrow$) & 32.90 \\
\hline\hline
\end{tabular}
}
\end{table*}

In absence of experimental results for most of the systems considered in this work, the validation of our findings largely depended upon agreement with the results of only a couple of experiments on Fe and Co substituting Mn and Ni respectively in Ni$_{2}$Mn$_{1.52}$Sb$_{0.48}$, a composition very close to the one considered here. In case of the system with Co substituting Ni, we observed a good agreement for the trends in magnetic moment and T$_{M}$ \cite{anayak2009}. For the system where Fe substitutes Mn, following discrepancies between our calculations and the experimental observations \cite{59sahoo2011} were found: (i) the magnetic moment of the system, decreases with Fe concentration as per our calculations, in complete disagreement with the trend observed in the experiment and (ii) although the variation of C$^{\prime}$, the most reliable predictor of T$_{M}$, with Fe concentration, suggesting a lowering of T$_{M}$, in agreement with the experimental findings, the trend in $\Delta E$, another predictor, and a useful one, as it can provide a quantitative estimate of T$_{M}$, implies the opposite. In this section, we attempt to find the origin of the discrepancy and resolve it. 
\begin{figure*}[t]
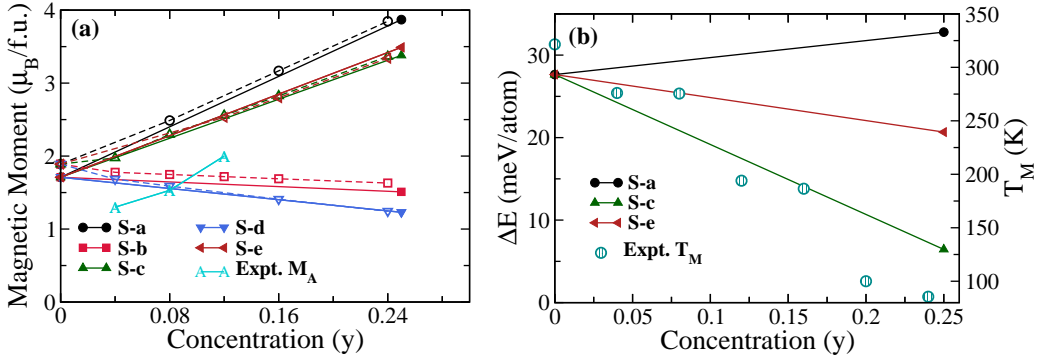

\centerline{\hfill
\psfig{file=fig6a.eps,width=0.345\textwidth}
\hspace{0.1cm}
\psfig{file=fig6b.eps,width=0.40\textwidth}
\hfill}
\caption{Variations in the (a) total magnetic moments with $y$ for various site configurations (Table \ref{table5}) for Ni$_{2}$Mn$_{1.50-y}$Fe$_{y}$Sb$_{0.50}$ system in their austenite phases calculated with VASP (closed symbols with solid line) and for  Ni$_{2}$Mn$_{1.52-y}$Fe$_{y}$Sb$_{0.48}$ calculated by SPRKKR (open symbols with dashed line) Experimental values of the total moment are added for comparison; (b)  total energy differences ($\Delta$E) between the austenite and martensite phases of Ni$_{2}$Mn$_{1.50-y}$Fe$_{y}$Sb$_{0.50}$ with $y$, for select site configurations. Experimental values of T$_{M}$ \cite{59sahoo2011} are given for comparing trends.}
\label{fig6}
\end{figure*}

\begin{figure}[htpb!]
\centerline{\hfill
\psfig{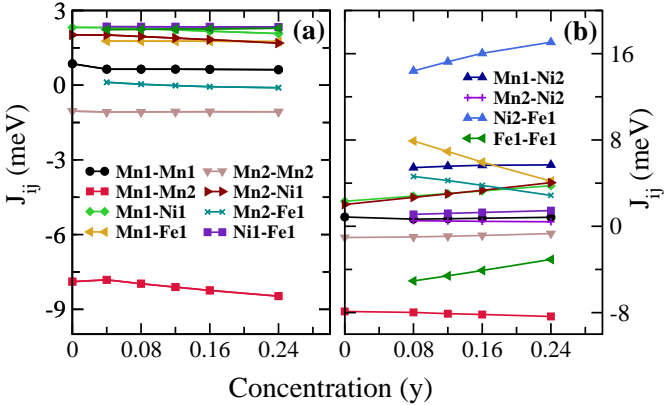}
\hfill}
\caption{The dependence of the inter-atomic magnetic exchange parameters in the first coordination shell (for different pairs of atoms) for Ni$_{2}$Mn$_{1.52-y}$Fe$_{y}$Sb$_{0.48}$ in their austenite phases for site configurations (a) $``$S-b$"$ and (b) $``$S-c$"$.}
\label{fig7}
\end{figure}

The possible origin of the discrepancy is the difference between the site ordering obtained in our calculation and the actual one generated during the experiment. The site ordering in an experimental sample depends on the thermal treatment. To resolve whether the site ordering is behind the observed discrepancy, we have done a detailed investigation considering all possible site occupancy of the constituent atoms in the Fe-doped Ni$_{2}$Mn$_{1.25}$Fe$_{0.25}$Sb$_{0.50}$ system. In Table~II of supplementary material we have listed  the possible site occupancies and magnetic configurations. The energies of the ground state magnetic configurations, for each pattern of site occupancy, are summarised in Table \ref{table5}. We, then, have calculated the magnetic moments as a function of Fe concentration between $0$ and $0.25$, for each site occupancy configuration of Table \ref{table5}. The results are shown in Figure \ref{fig6}(a). We find that the magnetic moment increases monotonically with Fe concentration, as observed in experiments, for all configurations except $``$S-b$"$ and $``$S-d$"$. It is to be noted that the configuration $``$S-b$"$ has the lowest energy for this compound and has been considered for calculations of physical properties throughout the paper. Since the configurations $``$S-a$"$, $``$S-c$"$, $``$S-e$"$ reproduce the experimentally observed trend of the magnetic moment, we next calculate the energy profiles of this system as a function of $(c/a)$ for these three configurations and compute $\Delta E$ in each case. The energy profiles are presented in Figure 7 of supplementary material and $\Delta$E are shown in Figure \ref{fig6}(b). In Figure \ref{fig6}(b), experimentally obtained T$_{M}$ are shown to find out the proximity of trends in variations of T$_{M}$ with one or more of the calculated $\Delta$E. We find that the configuration $``$S-c$"$, one where the substituent Fe atoms occupy the Ni sites while Ni atoms occupy the vacant Sb sites, provide the best agreement to the experiments, in terms of trends in magnetic moment of the austenite phase and the T$_{M}$. That this configuration provides the same trend in T$_{M}$ as the experiment, is further established when our calculated shear modulus C$^\prime$ is found to increase with Fe concentration (Table III, supplementary material). The calculated T$_{c}^{A}$ (Table III, supplementary material) provides further credence to $``$S-c$"$ being the site occupancy configuration realised in the experiment since its trend with Fe concentration also agrees to that in the experiment. The total energies of $``$S-b$"$ and $``$S-c$"$ are only 10 meV per atom apart. Thus, occurrence of the $``$S-c$"$ configuration during the heat treatment of the sample has substantial possibility. To conclude, the experimental results on Ni$_{2}$Mn$_{1.52-y}$Fe$_{y}$Sb$_{0.48}$ can be consistently interpreted by taking into consideration the role of site occupancies.

 In order to gain further insights into the connection between the site occupancies, large magnetic moment in the austenite phase and subsequently the MCE in this particular compound, we look at the variations in the magnetic exchange interactions as a function of $y$. The detailed comparison of the nearest neighbour inter-atomic magnetic exchange for $``$S-b$"$ and  $``$S-c$"$ configurations is  in Figure \ref{fig7}. We find that substantially large ferromagnetic interactions in $``$S-c$"$ configuration, the greatest being in the Ni2-Fe pair, makes the difference. The presence of large ferromagnetic interactions in the austenite phase produces a large moment, lends more stability to the austenite phase (T$_{M}$ reduces) and can be correlated with the large MCE observed experimentally. 


\section{Summary and Conclusions}
Ni-Mn based Heusler compounds have turned out to be promising materials for magneto-caloric applications. Substitution of the transition metals Ni and Mn by other $3d$ transition metals in ternary Ni-Mn based compounds, have been found to open up avenues for enhancing the magneto-caloric effects in these compounds. In this work, we have explored the potentials of Mn-excess, Sb-deficient Ni$_{2}$MnSb compounds, as magneto-caloric materials by substitution of Ni and Mn by $3d$ transition metals Fe, Co and Cu. Apart from being able to explain the trends of variations in quantities like the martensitic transformation temperature, magnetic transition temperature and the magnetic moments with compositions, observed in handful of experiments on this system, we have provided insights into the possibilities of significant magneto-caloric effects in this group of compounds; the ones which are yet to be synthesised. We found that the site occupancies of various atoms play an important role in the variations of the above mentioned physical quantities. The structural stabilities in these systems could be correlated to the magnetic exchange interactions and their variations. We predict that the compounds Ni$_{2}$Mn$_{1.5-y}$Co$_{y}$Sb$_{0.5}$ and Ni$_{2}$Mn$_{1.5-y}$Cu$_{y}$Sb$_{0.5}$; $y \sim 0.25$, can emerge as materials with large magneto-caloric effects. In conclusion, this work systematically explores the physics behind occurrence of magneto-caloric effect in substituted Ni-Mn-Sb compounds. The approach adopted and knowledge obtained from this work can be used to investigate a wider pool of materials, boosting the possibility to discover more materials with large MCE.

\section{Acknowledgement}
The authors greatfully acknowledge the Department of Science and Technology, India for the compuational facilities under Grant No. SR/FST/P-II/020/2009 and IIT Guwahati for the PARAM supercomputing facility.

\bibliographystyle{aip} 

\end{document}